\documentclass[12pt]{iopart}

\usepackage{iopams}  
\usepackage{graphicx} 
\usepackage{hyperref}

\def\fm#1{\ifmmode #1 \else $#1$\fi}
\def\ket#1{{%
  \ifmmode |\,#1\,\rangle \else $|\,#1\,\rangle$\fi}}
\def\bra#1{{%
  \ifmmode \langle\,#1\,| \else $\langle\,#1\,|$\fi}}

\def\nCa40{\fm{^{40}\mathrm{Ca}}\xspace}
\def\Ca40{\fm{\nCa40^{+}}\xspace}

\begin{document}

\title{Deterministic single-photon source from a single ion}

\author{H~G~Barros$^{1,2}$, A~Stute$^{1,2}$, T~E~Northup$^{1}$, C~Russo$^{1,2}$, P~O~Schmidt$^{1}$ and R~Blatt$^{1,2}$}

\address{$^{1}$ Institut f\"{u}r Experimentalphysik, Universit\"{a}t Innsbruck,
                6020 Innsbruck, Austria}
\address{$^{2}$ Institut f\"{u}r Quantenoptik und Quanteninformation (IQOQI),
                6020 Innsbruck, Austria}

\ead{helena.barros@uibk.ac.at}

\begin{abstract}
We realize a deterministic single-photon source from one and the same calcium ion interacting with a high-finesse optical cavity.  Photons are created in the cavity with efficiency $(88\pm17)\%$, a tenfold improvement over previous cavity-ion sources.  Results of the second-order correlation function are presented, demonstrating a high suppression of two-photon events limited only by background counts. The cavity photon pulse shape is obtained, with good agreement between experiment and simulation.  Moreover, theoretical analysis of the temporal evolution of the atomic populations provides relevant information about the dynamics of the process and opens the way to future investigations of a coherent atom-photon interface.    

\end{abstract}

\pacs{42.50.Pq, 32.80.Qk}

\maketitle

\section{Introduction}

Single photons represent an important resource in quantum information science \cite{Zoller:2005, Monroe:2002}, as basis elements in both linear optical quantum computing and quantum cryptography \cite{Knill:2001a, Raussendorf:2000, Gisin:2002} and as ``flying qubits'' travelling between the nodes of a quantum network \cite{Cirac:1997, Gheri:1998}.  The generation of single photons through the process of spontaneous decay from a single excited emitter has been demonstrated in diverse systems \cite{Grangier:2004}, for example in molecules \cite{Brunel:1999, Lounis:2000}, color centers in diamonds \cite{Brouri:2000, Kurtsiefer:2000}, quantum dots \cite{Santori:2001, Press:2007}, neutral atoms \cite{Darquie:2005} and ions \cite{Maunz:2007}.  Efficient collection of these photons presents a challenge, which has been addressed by coupling the emitter to a resonator in the framework of cavity quantum electrodynamics \cite{Berman:1994}. A more fundamental difficulty lies in the context of quantum networks, in which a coherent process is required in order to map quantum states between atoms and photons.  

The latter problem can be overcome when the coupling to the resonator is coherent, in which case we can exploit a vacuum-stimulated process to generate single photons within the cavity mode, as has been realized using trapped neutral atoms and ions \cite{Mckeever:2004, Hijlkema:2007, Keller:2004a}.  In the case of neutral atoms, the emitter is confined in an optical dipole potential within a high-finesse optical cavity.  The interaction between atom and cavity field is achieved via a Stimulated Raman Adiabatic Passage (STIRAP) process \cite{Bergmann:1998}; a coherent field applied between a ground and an excited state of the atom transfers the atom to a second ground state while creating a photon in the cavity.  Significant accomplishments include strong coupling between the atom and cavity \cite{Miller:2005}, generation of cavity photons with near-unit efficiency \cite{Mckeever:2004} and the ability to quantify photon-emission statistics for a single atom \cite{Hijlkema:2007}.  However, trap lifetimes of neutral atoms in cavities are currently limited to at most one minute \cite{Hijlkema:2007}.  In addition, the strength of the atom-cavity coupling varies from atom to atom due to the geometry of the dipole trap, and the precise control of atom number presents challenges. Ion-trap systems face greater obstacles in coupling strongly to a cavity, due to the necessity of integrating the cavity with the ion-trap structure.  In contrast to neutral atoms, however, ions offer the opportunity for extended trap lifetimes and known spatial localization within the cavity, including precise positioning with respect to the cavity standing-wave mode \cite{Guthohrlein:2001, Mundt:2002, Russo:2009}.  Furthermore, experiments with trapped ions have already demonstrated significant landmarks in the role of stationary qubits in quantum computation, including quantum logic gates and entanglement of up to eight ions \cite{Leibfried:2003a, Haffner:2005, Leibfried:2005, Benhelm:2008b}.  Single-photon generation has been achieved for a single ion in a cavity using the STIRAP process, which moreover enables tailored control of the photon's waveform \cite{Keller:2004a}.

Here we report on the realization of an efficient single-photon source using a single calcium ion trapped within a high-finesse optical cavity.  Our system shares some features with the original ion-trap single-photon experiment \cite{Keller:2004a}, including use of the same ion species and transition, although we employ a far-detuned Raman process rather than a STIRAP process.  However, while the previous realization generated intracavity photons with at most $(8.0 \pm 1.3 )\%$ efficiency, we are able to produce photons with near-unit efficiency, in $(88 \pm 17)\%$ of all attempts.  This improvement is due in part to a more favorable set of cavity parameters and is aided by the narrow linewidth of our classical drive field.  Rather than driving simultaneous transitions between multiple Zeeman sublevels, we address a specific transition, which we resolve via Raman spectroscopy \cite{Russo:2009}.  We use master-equation simulations of the eight-level ion and two cavity modes in order to guide our choice of parameters, with the goal of optimizing the photon generation efficiency.  Strong agreement of our simulations with experimental results confirms that we have developed a powerful model of our multilevel system, allowing us to investigate the dynamics of the photon generation process.

\begin{figure} [ht]
\begin{center}
\includegraphics{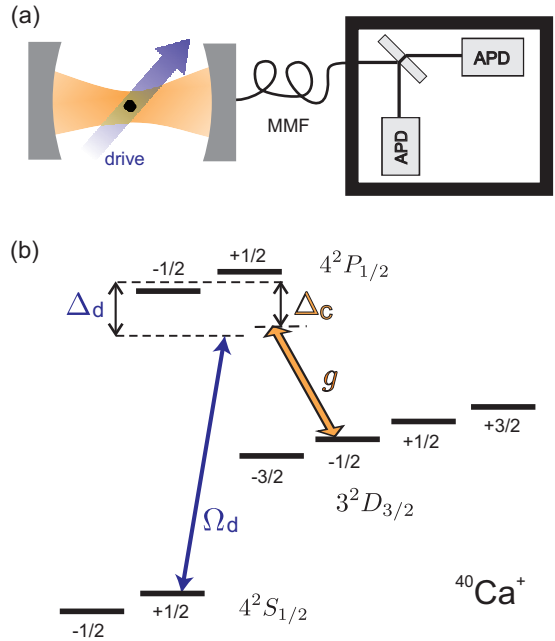}
\caption{Experimental setup. (a) Trapped ion surrounded by a high-finesse optical cavity and addressed by a drive laser. Photons produced in the cavity are sent through a multimode fibre (MMF) to the Hanbury Brown--Twiss setup, which consists of a beamsplitter and two avalanche photodiodes (APDs). (b) Relevant level scheme of the $^{40}$Ca$^+$ ion. A drive laser on the $S_{1/2}-P_{1/2}$ transition (Rabi frequency $\Omega_{d}$ and detuning $\Delta_{d}$) together with the cavity on the $P_{1/2}-D_{3/2}$ (coupling constant $g$ and detuning $\Delta_{c}$) perform a Raman transition between the Zeeman sublevels  $\ket{\mathrm{\textit{S}}_{1/2}, m = +1/2}$ and $\ket{\mathrm{\textit{D}}_{3/2}, m = -1/2}$. }
\label{fig:setup}
\end{center}
\end{figure}	

\section{Overview of experiment}

\subsection{Experimental apparatus}
A detailed description of the experimental setup can be found in \cite{Russo:2009}. In brief, we trap a single $^{40}$Ca$^+$ ion in a linear Paul trap situated in the center of a high-finesse optical cavity; the cavity is 2~cm long and has asymmetric mirror reflectivities. The ion and the cavity field interact via the atomic transition $\ket{\mathrm{\textit{P}}_{1/2}, m = +1/2}\fm{\leftrightarrow}\ket{\mathrm{\textit{D}}_{3/2}, m = -1/2}$ at 866~nm (cavity finesse of $70~000$) with a maximum coupling strength of $g_0 = 2\pi\times1.6$~MHz.  With a decay rate given by $2\kappa = 2\pi\times0.108$~MHz, the photons leave the cavity and are guided by a multimode fibre to a Hanbury Brown--Twiss (HBT) setup, as shown in Figure \ref{fig:setup}(a). This setup is composed of a 50/50 beamsplitter and two avalanche photodiodes (APDs) with measured quantum efficiencies of $(41\pm2)\%$ and $(42\pm2)\%$ and allows us to realize correlation measurements between photons.  A series of dichroic mirrors and optical filters eliminate stray light as well as transmission from the 785 nm laser used to stabilize the cavity length.

\subsection{Raman process}	
The interaction between ion and cavity is achieved via a vacuum-stimulated Raman transition between the \textit{S} and \textit{D} manifolds, in which a drive laser at 397~nm (linewidth $\approx30$~kHz) constitutes the first arm and the cavity represents the second arm of the transition, as shown in the level scheme of Figure \ref{fig:setup}(b). A magnetic field of $B = 0.2$~mT is applied to the ion perpendicular to the cavity axis in order to lift the Zeeman degeneracy. Both laser and cavity are detuned from the \textit{P}$_{1/2}$ manifold by $\Delta_{d}\approx\Delta_{c}\approx335$~MHz. The actual values of these detunings are set to fulfill one of the six possible Raman resonances between the two \textit{S} and four \textit{D} sublevels \cite{Russo:2009}. In this off-resonant scheme, we can write the effective Rabi frequency of this Raman transition and the effective atomic decay rate as 
	 $\Omega_{\mathrm{eff}}\approx g\cdot \Omega_d/2|\Delta_d|$ and
	 $\gamma_{\mathrm{eff}}\approx \gamma\cdot$($\Omega_d/2|\Delta_d|$)$^{2}$, respectively, where $\Omega_d$ is the Rabi frequency of the drive laser \cite{Dubin:2009}. 
	 With well-chosen values of $\Delta_d$ and $\Omega_d$, it is possible to achieve an intermediate regime of operation in which $\Omega_{\mathrm{eff}}\approx\gamma_{\mathrm{eff}}$. In this regime, the increased probability of realizing the Raman transition  when compared to the effective atomic spontaneous decay rate allows coherent processes to become more pronounced.

\subsection{Pulse sequence and data acquisition}
\label{pulse_sequence}
The experiment is realized in a pulsed fashion. An initial drive laser pulse of 120~$\mu$s generates a photon in the cavity, followed by a wait interval of 250~$\mu$s. This long wait interval is not necessary but allows us to resolve photon correlation data easily. As the \textit{D}$_{3/2}$ state is metastable, a laser pulse of 20~$\mu$s at 866~nm coupling the \textit{D}$_{3/2}$ and \textit{P}$_{1/2}$ manifolds is then applied to recycle the atomic population, in conjunction with a red-detuned beam for Doppler cooling on the $S_{1/2}-P_{1/2}$ transition.  Finally, a 20~$\mu$s optical pumping laser pulse at 397~nm between the $\ket{\mathrm{\textit{S}}_{1/2}, m = -1/2}$ and $\ket{\mathrm{\textit{P}}_{1/2}, m = +1/2}$ sublevels prepares the ion in the initial state $\ket{\mathrm{\textit{S}}_{1/2}, m = +1/2}$ with an efficiency $>98\%$; the recycling laser remains on during this stage. This 414.5~$\mu$s sequence is repeated 3500 times within each run, interleaved with intensity stabilization for the drive and recycling lasers; we obtain 4102 runs over the duration of three hours. Photon events are recorded in the two APDs with a resolution of 4~ps and an efficiency of $\eta_{\mathrm{det}} = (5.1\pm1.0)\%$, where this number represents the probability that a photon in the cavity will be detected at the APDs.  All data presented here were obtained with one and the same ion, its coupling to the mode of the cavity controlled precisely over the course of many hours.  Sudden episodes in which the ion becomes heated are observable as brief dropouts of the signal from the cavity and were discarded from the data obtained, comprising 174 runs (eight minutes of data).  In addition, 219 runs (ten minutes) were discarded due to the onset of a rapid drift in the laser frequencies of the experiment.

\begin{figure} [ht]
\begin{center}
\includegraphics[scale=0.5]{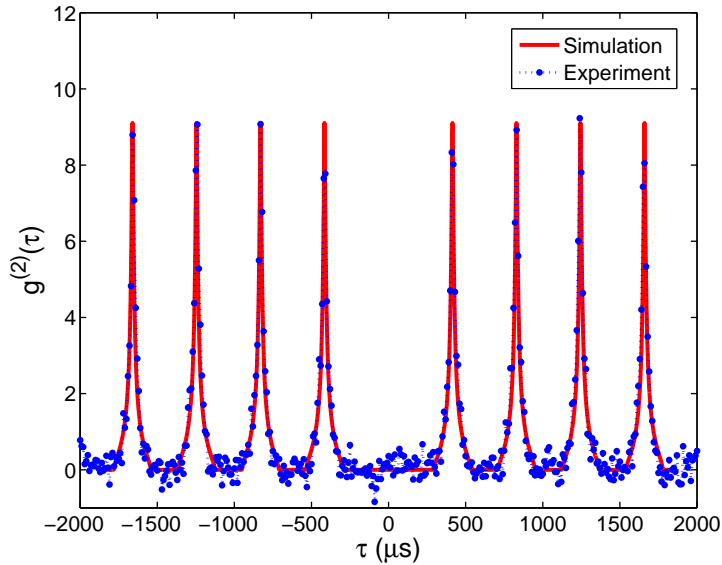}
\caption{Normalized second-order temporal correlation function of the single-photon source, with background subtracted. Experimental data (blue dotted line) is compared with a theoretical simulation (red straight line). The time-bin resolution is 1~$\mu$s, and the repetition period is 414.5~$\mu$s.}
\label{fig:g2exp}
\end{center}
\end{figure}

\section{Single-photon source statistics}
\label{data}

\subsection{Temporal correlation function and pulse shape}
The second-order temporal correlation function $g^2(\tau)$ is obtained from cross-correlations between photon arrival times at the two detectors as a function of time delay $\tau$, normalized to the mean intensity; the results are shown in Figure \ref{fig:g2exp}. The temporal structure of $g^2(\tau)$ reveals the characteristics of a pulsed source of light: the individual peaks are separated by the drive pulse period of 414.5~$\mu$s, and their width is given by the convolution of the waveform of two single pulses. The defining characteristic of a single-photon source is evident in the absence of a peak at $\tau=0$, which reflects the probability of detecting two photons within the same drive sequence.  We observe a total of $150 \pm 150$  background-subtracted counts in a window of $\pm 207.25~\mu$s around $\tau=0$.   Suppression is thus limited entirely by background counts, which consist solely of APD dark counts. In comparison, we detect $581 000 \pm 2000$ single photons over the measurement period. Background subtraction is accomplished as described in \cite{Keller:2004}.  We introduce an artificial dead time of $2.5~\mu$s to avoid spurious counts due to after-pulsing of our detectors, which occurs with probability 1.1$\%$ within this window. 
In addition, we observe a narrow peak in cross-correlations at $\tau = 125$~ns corresponding to a photon emitted by one APD during a detection event, then reflected back from the cavity output mirror to the other detector.  These reflected photons are confined to a 20~ns window, which is removed from the data \cite{Russo:2008}.
The single-photon nature of our source is due to the fact that exactly one ion is trapped within an optical cavity. After emitting a photon into the cavity mode, the ion occupies the metastable $\mathrm{\textit{D}}_{3/2}$ state with a one-second lifetime, ensuring no further coupling to the drive beam. 

\begin{figure} [ht]
\begin{center}
\includegraphics[scale=0.5]{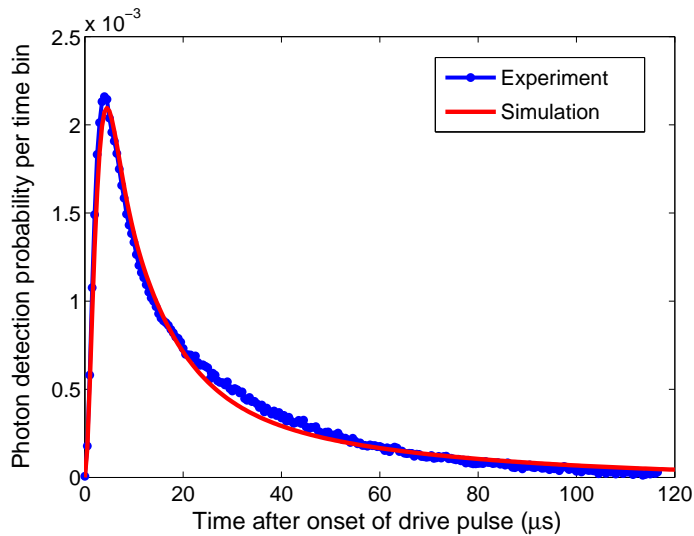}
\caption{Photon pulse shape in the time window of a drive pulse. For each 500~ns time bin, we plot the probability to detect a photon (blue dotted line). Simulations are superimposed (red straight line).}
\label{fig:photonwavepacket}
\end{center}
\end{figure}

In order to evaluate the pulse shape of the photon exiting the cavity, we generate a histogram of the time interval between the start of the drive pulse and detection of a photon.  This histogram is then normalized by the number of trials, and the resulting probability distribution of photon arrival times per time bin is shown in Figure \ref{fig:photonwavepacket}.  The total area of $\eta_{\mathrm{exp}} = 0.045$ under the curve thus represents the probability to detect a photon in the course of a single trial.  From our previous measurement of the detection efficiency, we calculate a photon creation efficiency of $\eta_{c} = \eta_{\mathrm{exp}}/\eta_{\mathrm{det}} = (88\pm17)\%$.

\subsection{Simulations}
\label{simulations}

Also shown in Figure \ref{fig:photonwavepacket} is the result of a master-equation simulation of the eight-level $^{40}\mathrm{Ca}^{+}$ ion and two orthogonal cavity modes, each with a truncated basis of three Fock states.  Linewidths of the atom, cavity and drive field are included as decoherence channels in the Liouvillian \cite{Carmichael:1993}.  We perform independent calibration measurements for all parameters, including magnetic field and Rabi frequency, detuning and linewidth of the drive field.  In order to find the best agreement with the data, we then adjust three of these parameter values within the error range of their measurement.  Specifically, we use a Rabi frequency of the drive field $\Omega_{d} = 2\pi \times 30$~MHz, given our calibrated value of $2\pi \times (40 \pm 10)$~MHz.  Second, the cavity and drive laser frequencies are initially set to meet the Raman resonance condition, but over the course of several hours of data acquisition, their relative detuning drifts by a few hundreds of kHz.  It is not realistic to reproduce this drift in our simulations, and we instead use an average detuning value of $2\pi \times 60$~kHz.  Finally, the measured photon detection efficiency of 0.045 is taken into account in the simulation result by appropriate scaling of the amplitude. This scaling allows us to infer an output path efficiency of 6.1$\%$, consistent with $\eta_{\mathrm{det}} = (5.1\pm1.0)\%$ (section~\ref{pulse_sequence}).  The strong agreement that we are able to obtain with experiment suggests that our simulations provide a realistic model of the complex dynamics of our atom-cavity system, which in turn allows us to understand the relative processes of cavity-stimulated and spontaneous photon emission.  For the parameters shown in Figure \ref{fig:photonwavepacket}, we extract a photon creation efficiency of $74\%$, consistent with the experimentally determined value $\eta_{c}= (88\pm17)\%$.
	 
\begin{figure} [ht]
\begin{center}
\includegraphics[scale=0.5]{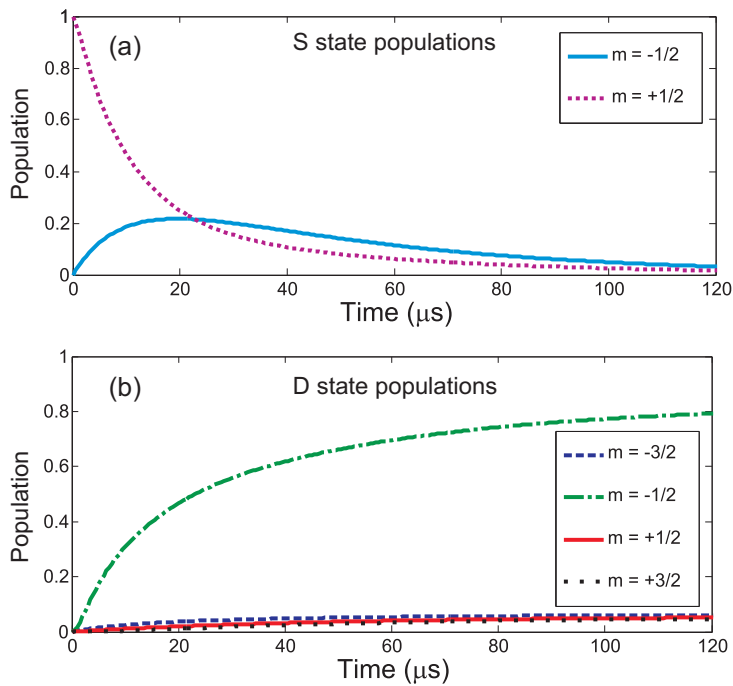}
\caption{Temporal evolution of atomic populations in the (a) \textit{S}$_{1/2}$ manifold, and (b) \textit{D}$_{3/2}$ manifold, in the time window of a drive pulse. }
\label{fig:pops}
\end{center}
\end{figure}

Simulations also provide important information about the temporal evolution of the atomic populations.  Before the photon generation sequence begins, we initialize the ion in the state $\ket{\mathrm{\textit{S}}_{1/2}, m = +1/2}$.  We then introduce the drive pulse, which is tuned to Raman resonance with the state $\ket{\mathrm{\textit{D}}_{3/2}, m = -1/2}$, in order to begin population transfer between the two states.  However, due to spontaneous scattering from the excited $P$ manifold to the $S$ and $D$ manifolds, and to a lesser extent to off-resonant Raman transfer, population also accumulates in the remaining four $S$ and $D$ sublevels.  This process is shown in Figure \ref{fig:pops}, which plots the $S$ and $D$ populations as a function of time during the interval when the drive pulse is active.  In Figure \ref{fig:pops}(a), the population in the initial state $\ket{\mathrm{\textit{S}}_{1/2}, m = +1/2}$ is seen to decrease with time as expected.  In addition, a small fraction of population is transfered to the $\ket{\mathrm{\textit{S}}_{1/2}, m = -1/2}$ state, reaching a maximum within approximately 20~$\mu$s, after which the state slowly depopulates. This population results from a Raman scattering process in which the excited state \textit{P}$_{1/2}$ is off-resonantly driven. 
Rayleigh scattering back to the initial $\ket{\mathrm{\textit{S}}_{1/2}, m = +1/2}$ state is also possible, so that the intended process of Raman transfer to the $D$ manifold may only occur after multiple scattering events.

In Figure \ref{fig:pops}(b), the corresponding accumulation of population in the four \textit{D} sublevels is plotted. (Note that the fractional population in the two $P$ sublevels remains negligible, that is, on the order of $10^{-4}$.)  We emphasize that the laser-cavity Raman transition is indeed the dominant process, and the population of the $\ket{\mathrm{\textit{D}}_{3/2}, m = -1/2}$ state reaches approximately $80\%$ at the end of the pulse interval.  The populations of the three remaining \textit{D} sublevels are primarily due to spontaneous decay from the upper state \textit{P}; also, excited state decay to the target state $\ket{\mathrm{\textit{D}}_{3/2}, m = -1/2}$ is responsible for $6\%$ of its final population, in addition to the $74\%$ reported earlier, attributed to the photon generation process.

These dynamics are highly sensitive to input parameters and have thus guided our selection of appropriate experimental values as well as our understanding of the limits of our apparatus.  For example, we have seen that the narrow linewidth of our drive field, a recent technical upgrade, results in a significant improvement in photon generation efficiency.  In addition, we have selected the Rabi frequency and detuning of our drive field in order to work in a regime of high photon generation efficiency.  While in principle, a smaller effective Rabi frequency for the Raman transition would further increase the efficiency, the process would in this case become too sensitive to previously mentioned frequency drifts in the laboratory.
	
\section{Conclusion and outlook}	

We have demonstrated a highly efficient ion-cavity single-photon source and characterized its output pulse shape and dark-count-limited suppression of two photon events.  Such a source offers the prospect for coherent state transfer between atoms and photons within a quantum network.  However, one must consider the obstacles that realistic experimental parameters present for this transfer process.  The simulations of section \ref{simulations} represent an important step in this direction, as they allow us to analyze processes which would destroy coherence during photon generation.  Consider, for example, the Raman and Rayleigh scattering processes which may occur between the $P$ and $S$ manifolds.  The results of \cite{Ozeri:2005, Ozeri:2007} have shown that while Raman scattering events introduce decoherence to the system, coherence may be preserved during Rayleigh scattering, which returns the atom to its initial state.  From this perspective, it is interesting to evaluate the percentage of photons generated in the cavity without Raman scattering.  We can explore this situation by modifying our simulation to include an artificial, ninth atomic level, accessed only through rapid decay from the state $\ket{\mathrm{\textit{S}}_{1/2}, m = -1/2}$; the new level thus functions as a dark state which collects any population that undergoes Raman scattering.  By comparing the modified simulation to the original, we find that $70.3\%$ of cavity photons belong to this target group, that is, most photon generation events do not include a Raman scattering process.

Moreover, we anticipate that we can increase this fraction by truncating the length of the drive pulse (for the same Rabi frequency).  This is suggested by Figure \ref{fig:pops}(a): the $\ket{\mathrm{\textit{S}}_{1/2}, m = -1/2}$ population accumulates slowly, and its long tail is due to events where the ion must wait for another scattering event before it can undergo Raman transfer from the $\ket{\mathrm{\textit{S}}_{1/2}, m = +1/2}$ state.  Thus, by eliminating later events, we suppress Raman scattering, at the cost of a reduction in photon generation efficiency.  To confirm this, we truncate the drive pulse in our simulation by a factor of ten, so that it has a length of 12~$\mu$s. In this case, while the simulated photon generation efficiency drops from $74\%$ to $30\%$, we find that $98.3\%$ of the cavity photons are created without Raman scattering.  Although our current experiments start with a pure state, where no coherence is stored in the ion, we find these results promising for future efforts in which the ion is prepared in a superposition state, with the goal of deterministic state transfer to a photon. 

\section{Acknowledgments}

 This work has been partially supported by the Austrian Science Fund (SFB 15), by the European Commission (QUEST network, HPRNCT-2000-00121, QUBITS network, IST-1999-13021, SCALA Integrated Project, Contract No. 015714), and by the ``Institut f\"ur Quanteninformation GmbH.'' C. Russo acknowledges support from the Funda\c{c}\~{a}o para a Ci\^{e}ncia e a Tecnologia -- SFRH/BD/6208/2001, and A. Stute acknowledges support from the Studienstiftung des deutschen Volkes.


\end{document}